\documentclass{ws-ijmpd}
\usepackage[super]{cite}
\usepackage{xcolor}
\usepackage[verbose,hypertexnames=false]{hyperref}
\hypersetup{colorlinks=false,allbordercolors=blue,pdfborderstyle={/S/U/W 1}}
\usepackage{graphicx}
\usepackage{amsbsy,amsmath}
\usepackage{amsfonts}
\usepackage{mathtools}
\usepackage{empheq}
\usepackage{microtype}
\usepackage[macrosonly]{chet}
\usepackage{xcolor}
\definecolor{darkblue}{rgb}{0.1,0.1,.7}
\definecolor{darkgreen}{rgb}{0,.5,0}
\usepackage{hyperref}
\usepackage{booktabs}
\usepackage{array}
\usepackage{mdframed}
\newmdenv[linecolor=black,linewidth=1pt,roundcorner=6pt,backgroundcolor=gray!10,
  skipabove=10pt,skipbelow=10pt,innerleftmargin=10pt,innerrightmargin=10pt,
  innertopmargin=8pt,innerbottommargin=8pt]{mybox}
\renewcommand{\wsdraftnote}{}

\def\bz{{\overline{z}}}

\def\sc{{\hat{c}}}

\def\cI{{\cal I}}
\def\cZ{{\cal Z}}
\def\ze{\zeta}
\def\bze{\bar{\zeta}}

\newcommand{\cO}{{\cal O}}

\DeclareMathOperator{\Tr}{Tr}

\def\be{\begin{equation}}
\def\ee{\end{equation}}
\def\bea{\begin{eqnarray}}
\def\eea{\end{eqnarray}}

\def\Tr{{\rm Tr}}

\def\a{\alpha}
\def\b{\beta}

\def\G{\Gamma}

\def\m{\mu}

\def\z{\bar{z}}

\def\cQ{{\cal Q}}
\def\cD{{\cal D}}

\def\cF{{\cal F}}
\def\cZ{{\cal Z}}
\def\cI{{\cal I}}
\def\ze{\zeta}
\def\bze{\bar{\zeta}}

\begin{document}

\title{A Thermal Representation for Conformal Ladder Integrals%
\footnote{Work with Manthos Karydas, Songyuan Li and Matthieu Vilatte.}}

\author{Anastasios C. Petkou}

\address{Laboratory of Theoretical Physics\\
Aristotle University of Thessaloniki, Greece\\
petkou@auth.gr}
\maketitle
\flushbottom

\begin{abstract}
I discuss the recently discovered representation of conformal four-point ladder integrals in terms of the thermal free energy of free massive scalar fields. These integrals satisfy a novel second-order differential equation in even dimensions $D$ at arbitrary loop order $L$.  I also present a simple derivation of the all-loop resummation of conformal ladder integrals for arbitrary $D$. Possible connections to the thermal bootstrap, multiloop calculations, integrability, AdS/CFT and string theory are briefly discussed. This is a proceedings contribution to the Athens Workshop in Theoretical Physics: 10th Anniversary, held at the National and Kapodistrian University of Athens on December 17--19 2025.
\end{abstract}


\section{Introduction}
\label{sec:intro}

Four-point functions are quantities of prime interest in conformal field theories (CFTs), as their knowledge leads—under general assumptions—to the complete solution of the theory. Beyond pure symmetry arguments, they can also be calculated perturbatively using Feynman integral methods. The first truly conformal technique for $n$-point integrals with $n\geq 3$ was presented in Symanzik's work \cite{Symanzik:1972wj}, applied to explicit CFT calculations some twenty years later \cite{Petkou:1994ad,Hoffmann:2000mx,Hoffmann:2000tr}, and more recently rediscovered and refined in the context of AdS/CFT \cite{Dolan:2000ut}. In a related development, Mellin transform techniques were used to evaluate multiloop four-point ``ladder'' integrals \cite{Usyukina:1992jd,Usyukina:1993ch}; their connection to conformal symmetry was emphasised in \cite{Broadhurst:1993ib,Broadhurst:2010ds}. Two highlights of subsequent progress are the connection to single-valued polylogarithms \cite{Brown:2004ugm,Schnetz:2013hqa} and to integrability \cite{Isaev:2003tk,Derkachov:2021ufp}.

In this work we discuss a novel relationship between a class of conformal four-point integrals and thermal one-point functions of higher-spin composite operators in massive free QFTs—the \textit{thermal representation of conformal ladder integrals}. Our observation originates in \cite{Iliesiu:2018fao, Petkou:2018ynm}, building on preliminary results in \cite{Petkou:2021zhg,Karydas:2023ufs}, and is fully developed in \cite{Karydas:2025tfs}.

\section{Conformal Ladder Integrals in $D=2k+2$ Dimensions}
\label{sec: conf ladder integrals}

We consider the following class of $L$-loop  ladder integrals in even dimensions $D=2k+2$, $k=0,1,2,\ldots$:
\be
\label{ILk_def}
g^{2L}I_L^k(x_1,x_2,x_3,x_4)=\prod_{n=1}^{L}\left[\int\frac{d^{2k+2}y_n}{\pi^{k+1}}
\frac{g^2\Gamma(k)}{|y_{n-1,n}|^{2k}|x_2-y_n|^2|x_4-y_n|^2}\right]\frac{1}{|y_L-x_3|^{2k}}\,,
\ee
where $y_0=x_1$, $y_{ij}=y_i-y_j$, $g^2$ is a dimensionless loop-counting parameter and $\Gamma(k)$ is introduced for later convenience. For $L=0$ we have $I_0^k= 1/|x_1 - x_3|^{2k}$. These yield conformally covariant four-point functions with weights $k$ and $L$ in the $(1$-$3)$ and $(2$-$4)$ channels respectively. Using conformal symmetry to fix $x_1\mapsto\ze$, $x_2\mapsto\infty$, $x_3\mapsto 1$, $x_4\mapsto 0$ one has
\be
\label{ILk_lim}
\cI_L^k(\ze,\bze)\equiv\lim_{\{x_i\}\mapsto\{\ze,\infty,1,0\}}\left[x_{24}^{2L}I_L^k\right]=\frac{1}{|1-\ze|^{2k}}\Phi_L^k(\ze,\bze)\,.
\ee
\begin{figure}[h!]
    \centering    \includegraphics[width=0.9\linewidth]{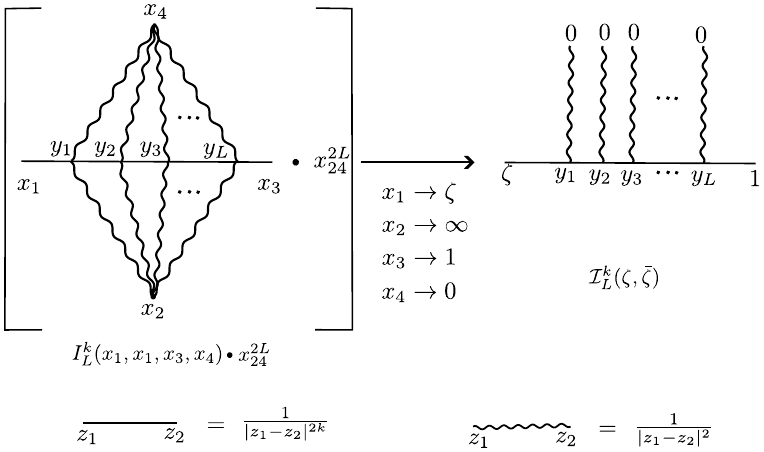}
    \caption{Graphical representation of $\mathcal{I}^{k}_{L}(\z, \bar{\z})$}
    \label{fig: conf integrals}
\end{figure}
The singular limit $k=0$ yields $\cI_L^0(\ze,\bze)$, which correspond to conformal integrals in $D=2$ dimensions and have been discussed in Appendix A of \cite{Karydas:2025tfs}. The functions $\Phi_L^k(\ze,\bze)$ encode all relevant dynamical information.  Using Isaev's integral representation \cite{Isaev:2003tk} with the change of variables $\ze'=\ze e^{-t}$ we obtain the following result \cite{Karydas:2025tfs}
\be
\label{ILk_our1}
\Phi_L^k(\ze,\bze)=\frac{1}{\cD_0^k(\ze,\bze)}\frac{1}{L!(L-1)!}
\int_0^{|\ze|}\frac{d|\ze'|}{|\ze'|}2\ln|\ze'|\,(\ln^2|\ze'|-\ln^2|\ze|)^{L-1}\cD_0^k(\ze',\bze')\,,
\ee
where
\be
\label{D0k}
\cD_0^k(\ze,\bze)=\Gamma(k)\frac{(\ze-\bze)^k}{|1-\ze|^{2k}}\,,\quad k=1,2,3,\ldots\,.
\ee
The representation (\ref{ILk_our1}), though a straightforward extension of \cite{Isaev:2003tk}, reveals a remarkable connection to thermal free energies as we show below.

\section{Thermal Free Energies of Massive Free Scalars from a Parent Quantum Mechanical System}
\label{sec: QM model}

\subsection{The parent quantum mechanical system}

The parent system consists of two harmonic oscillators with Hamiltonian
\be
\label{BasicHamiltonian}
\hat{H}=\tfrac{1}{2}\hat{p}^2_1+\tfrac{1}{2}\hat{p}^2_2+\tfrac{1}{2}m^2(\hat{x}^2_1+\hat{x}^2_2)
+i\mu\!\left(\hat{p}_2\hat{x}_1-\hat{p}_1\hat{x}_2\right)\,,
\ee
with unit mass, common frequency $m$ and $\m$ an imaginary twisting (chemical potential). Using that standard quantisation  one can calculate the grand canonical partition function
\begin{align}
\label{Z0def}
\cZ_0(z,\bar{z})&=\Tr_{{\cal H}_{1,2}}\!\left[e^{-\beta(\hat{H}_0+m^2\hat{\cO})}e^{-i\beta\mu\hat{\cQ}}\right]
=e^{\ln|z|-\ln(1-z)-\ln(1-\bz)}\,,
\end{align}
 and the free energy ${\cal F}_0$
\be
\label{defF}
\ln\cZ_0(z,\bz)=-\beta\cF_0(z,\bz)=\ln|z|-\ln(1-z)-\ln(1-\bz)\,,
\ee
with $z=e^{-\beta m-i\beta\mu}$.
Two key differential operators are
\begin{align}
\label{opD}
\hat{\bf D}_z&=\tfrac{1}{\beta^2}\tfrac{\partial}{\partial m^2}=\tfrac{1}{2\ln|z|}(z\partial_z+\bz\partial_\bz)\,,\\
\label{opL}
\hat{\,\bf L}_z&=\tfrac{i}{\beta}\tfrac{\partial}{\partial\mu}=(z\partial_z-\bz\partial_\bz)\,,
\end{align}
which commute, $[\hat{\bf D}_z,\hat{\,\bf L}_z]=0$ and give the Laplacian in $m,\mu$ variables as
\be
\label{Laplacian}
\hat{\bf\Delta}_z=\tfrac{\partial^2}{\partial m^2}+\tfrac{\partial^2}{\partial\mu^2}
=4\b^2 z\bz\partial_z\partial_{\bz}\,.
\ee
Acting on $\ln\cZ_0$ with $\hat{\bf D}_z$ and $\hat{\,\bf L}_z$ gives the thermal one-point functions of $\hat{\cO}=\tfrac{1}{2}(\hat{x}_1^2+\hat{x}_2^2)$ and $\hat{\cQ}=\hat{p}_2\hat{x}_1-\hat{p}_1\hat{x}_2$ as
\begin{align}
\label{eq: Oaverage}
\langle\hat{\cal O}(z,\bz)\rangle_0&=-\beta\hat{\bf D}_z*\ln \cZ_0(z,\bz)=\frac{1}{2m}\frac{1-|z|^2}{|1-z|^2}\,,\\
\label{eq:Qaverage}
\langle\hat{\cal Q}(z,\bz)\rangle_0 &=\hat{\,\bf L}_z*\ln \cZ_0(z,\bz)=\frac{z-\bz}{|1-z|^2}\,.\end{align}
Interestingly, these are related to Poisson kernels: $\langle\hat{\cal O}(z,\bz)\rangle_0$ and $\langle\hat{\cal Q}(z,\bz)\rangle_0$ are harmonic functions on the open unit disk  and on the upper half plane respectively, that take constant values on their corresponding boundaries. 

\subsection{Ideal relativistic gases and thermal free energies}

Viewing $\cZ_0$ as the partition function of a pair of ``photons'' with imaginary chemical potential, the free energy of an ideal relativistic photon gas in $d=2L+1$ dimensions is obtained by placing the system in a spatial box with volume $V_{2L}=\ell^{2L}$ and integrating over the one-particle density of states 
\be
\rho_L(\omega;m;\a^2)=\frac{2\a^{2L}\b^{2L}}{(L-1)!}\omega(\omega^2-m^2)^{L-1}\,,
\ee
where $\a^2=\ell^2/(4\pi\beta^2)$ is a geometric parameter:
\begin{align}
\label{Zdexpl}
\ln\cZ_L(z,\bz;\a^2)&=-\frac{\a^{2L}}{(L-1)!}\int_0^{|z|}\frac{d|z'|}{|z'|}2\ln|z'|\,
(\ln^2|z'|-\ln^2|z|)^{L-1}\ln\cZ_0(z',\bz')\,.
\end{align}
Setting $\ln\cZ_L(z,\bz;\a^2)=\a^{2L}\ln\cZ_L(z,\bz)$ the explicit result reads \cite{Petkou:2021zhg}
\begin{mybox}
\be
\label{ZLresult}
\ln\cZ_L(z,\bz)=\frac{(-1)^L L!(2\ln|z|)^{2L+1}}{2(2L+1)!}
+\sum_{n=0}^{L}\frac{(2L-n)!(-2\ln|z|)^n}{(L-n)!\,n!}\,2\Re\bigl[Li_{2L+1-n}(z)\bigr]\,,
\ee
\end{mybox}
where $Li_s(z)$ are standard polylogarithms.  Notice that for $m=\m=0$ we have
 $\ln\cZ_L(1,1)=\frac{(2L)!}{L!}2\zeta(2L+1)$. 
The result (\ref{ZLresult})  coincides with the logarithm of the thermal partition function of a free massive complex scalar \cite{Petkou:2021zhg} with Euclidean action
\be
\label{SE}
{\cal S}_L(\b;m,\m)=\int_0^\beta\!d\tau\int\!d^{2L}\vec{x}\,\left[|(\partial_\tau-i\m)\phi|^2+|\vec{\partial}\phi|^2+m^2|\phi|^2\right].
\ee
Acting with $\hat{\bf D}_z$ on $\ln\cZ_L$ not only gives the thermal one-point function $\langle\cO\rangle_L$, but simultaneously yields $\ln\cZ_{L-1}$. A similar result holds for the action of $\hat{\bf L}_z$:
\begin{align}
\label{Oaverage}
&\langle\cO(z,\bz)\rangle_L=-\beta\hat{\bf D}_z*\ln\cZ_L(z,\bz)=\b\ln\cZ_{L-1}(z,\bz)\,,\\
\label{Qaverage}
&\langle\cQ(z,\bz)\rangle_L=\hat{\,\bf L}_z*\ln\cZ_L(z,\bz)=-\hat{\bf D}_z*\langle\cQ\rangle_{L+1}\,.
\end{align}

\subsection{The iterated integral representation of the thermal free energy}

The free energies above admit the iterated integral representation
\begin{mybox}
\be
\label{ZL}
\ln\cZ_L(z,\bz)=(-1)^L\left\{{\bf\text{ord}}\prod_{i=1}^{L}\right\}
\left[\int_0^{|z_{i+1}|}\frac{d|z_i|}{|z_i|}2\ln|z_i|\right]\ln\cZ_0(z_1,\bz_1)\,,
\ee
\end{mybox}
where the ordered product denotes integration performed from left to right  with $0\leq|z_1|\leq\cdots\leq|z_L|\leq|z|$. This coincides with the class of iterated integrals giving rise to single-valued polylogarithms \cite{Brown:2004ugm,Schnetz:2013hqa}.

Defining the integral operator
\be
\check{\!\mathbf{d}}_{z;z'}=\int_0^{|z|}\frac{d|z'|}{|z'|}2\ln|z'|\,,
\ee
with  properties $\check{\!\mathbf{d}}_{z;z'}*\hat{\bf D}_{z'}=\hat{\bf D}_z*\check{\!\mathbf{d}}_{z;z'}=1$ and $[\check{\mathbf{d}}_{z;z'},\hat{\,\bf L}_{z'}]=0$, we can  write complactly
\be
\label{ZLd}
\ln\cZ_L(z,\bz)=[-\,\check{\!\mathbf{d}}]_{z;z'}^L*\ln\cZ_0(z',\bz')\,.
\ee
More formally, we may consider the differential operators $\hat{\bf D}_z$ and $\hat{\,\bf L}_z$ as a set of orthogonal vectors on the two-dimensional Euclidean space with complex coordinates $z,\bar{z}$ and metric $ds^2=dzd\bar{z}$. The corresponding dual one-forms are 
\be
\label{1forms}
\check{\bf D}_z=\ln|z|\left(\frac{dz}{z}+\frac{d\bar{z}}{\bar{z}}\right)\,,\,\,\,\,\check{\,\bf L}_z=\frac{1}{2}\left(\frac{dz}{z}-\frac{d\bar{z}}{\bar{z}}\right)\,,
\ee
such that the inner products are
\be
\label{InnerProducts}
\langle \check{\bf D}_z,\hat{\bf D}_z\rangle=1\,,\,\,\langle \check{\bf D}_z,\hat{\,\bf L}_z\rangle=0\,,\,\,\,\langle \check{\,\bf L}_z,\hat{\bf D}_z\rangle=0\,,\,\,\langle \check{\,\bf L}_z,\hat{\,\bf L}_z\rangle=1\,.
\ee
\subsection{Thermal representation for massless two-point functions}

One can verify that
\be
\label{ellOp}
(z-\bz)^k\!\left(\tfrac{1}{z-\bar{z}}\hat{\,\bf L}_z\right)^k\!*\ln\cZ_0(z,\bz)=\cD_0^k(z,\bz)=\Gamma(k)\frac{(z-\bz)^k}{|1-z|^{2k}}\,,
\ee
so massless two-point functions in $D=2k+2$ dimensions are thermal cumulants of the parent system. We also find that the operator $\hat{\,\bf L}_z^{(k)}\equiv\hat{\,\bf L}_z-k\tfrac{z+\bz}{z-\bz}$ raises the depth parameter $k\mapsto k+1$,  and the dimension \ $D\mapsto D+2$ as
\be
\label{Lzk}
\hat{\,\bf L}_z^{(k)}*\cD_0^k(z,\bz)=\cD_0^{k+1}(z,\bz)\,,\,\,\,k=0,1,2,3,... \, .
\ee
It also commutes with $\check{\bf d}_{z;z'}$ as
\be
\label{Lzkd}
\check{\!\mathbf{d}}_{z;z'}*\hat{\,\bf L}^{(k)}_{z'}*f(z')= \hat{\,\bf L}^{(k)}_z*\check{\!\mathbf{d}}_{z;z'}*f(z')\,.
\ee
The above motivate us to write\footnote{Note the difference between the operator $\hat{\,\bf L}_z^{(k)}$ and the exponentiation $[\hat{\,\bf{L}}_z]^k$.}
\be
\label{LzK_def}
\left\{\text{ord}\prod_{n=0}^{k-1}\right\}\left[\hat{\,\bf L}_z-n\frac{z+\bz}{z-\bz}\right]*\ln\cZ_0(z,\bz)\equiv [\hat{\,\bf{L}}_z]^k*\ln\cZ_0(z,\bz)=\cD_0^k(z,\bz)\,.
\ee
As in \eqref{ZL} we have used again the ordered product symbol to indicate that $n$ increases from right to left. We further notice that $\hat{\,\bf L}_z^{(k)}$ satisfies
$[\hat{\,\bf L}_z^{(k)},\hat{\bf D}_z]=0$. We summarize in Tables \ref{tab:my_label_1}, \ref{tab:my_label_2}  and \ref{tab:my_label_3}  the definitions and properties of the various operators defined above.

\begin{table}[h!]
    \centering
    \renewcommand{\arraystretch}{1.3}
    \begin{tabular}{>{$}l<{$} >{$}l<{$} >{$}l<{$}}
    \toprule
    \ln\cZ_0(z,\bz)  &\equiv & \cD_0^0(z,\bz) = \ln|z| - \ln(1-z) - \ln(1-\bz) \\
    \ln\cZ_L(z,\bz)  &\equiv& \cD_L^0(z,\bz) \\
    \cD_0^k(z,\bz)  & =& \dfrac{\Gamma(k)(z-\bz)^k}{|1-z|^{2k}},\quad k\geq 1 \\
    \bottomrule
    \end{tabular}
    \caption{Summary of definitions}
    \label{tab:my_label_1}
\end{table}

\begin{table}[h!]
    \centering
    \renewcommand{\arraystretch}{1.3}
    \begin{tabular}{>{$}l<{$} >{$}l<{$} >{$}l<{$}}
    \toprule
    \hat{\bf D}_z & =& \tfrac{1}{\beta^2}\tfrac{\partial}{\partial m^2}
      = \tfrac{1}{2\ln|z|}(z\partial_z+\bz\partial_\bz) \\[4pt]
    \hat{\,\bf L}_z & = &\tfrac{i}{\beta}\tfrac{\partial}{\partial \mu}
      = (z\partial_z-\bz\partial_\bz) \\[4pt]
    \hat{\,\bf L}_z^{(k)} & = &\hat{\,\bf L}_z - k \tfrac{z+\bz}{z-\bz} \\[4pt]
    [\hat{\,\bf L}_z]^k & =& (z-\bz)^k \Big(\tfrac{1}{z-\bz}\hat{\,\bf L}_z\Big)^k \\[4pt]
    \check{\!\mathbf{d}}_{z;z'} & = &\int_0^{|z|}\frac{d|z'|}{|z'|}\,2\ln|z'| \\
    \bottomrule
    \end{tabular}
    \caption{Differential and integral operators.}
    \label{tab:my_label_2}
\end{table}

\begin{table}[h!]
    \centering
    \renewcommand{\arraystretch}{1.3}
    \begin{tabular}{>{$}l<{$} >{$}l<{$} >{$}l<{$}}
    \toprule
    \ln\cZ_L(z,\bz) & =& [-\,\check{\!\mathbf{d}}]_{z;z'}^L * \ln \cZ_0(z',\bz') \\[4pt]
    \hat{\bf D}_z * \ln \cZ_L(z,\bz) & = &\ln \cZ_{L-1}(z,\bz)
       = -\tfrac{1}{\beta}\langle \cO(z,\bz)\rangle_L \\[4pt]
    \hat{\,\bf L}_z * \ln \cZ_L(z,\bz) & = &\langle\cQ(z,\bz)\rangle_L \\[4pt]
    \check{\!\mathbf{d}}_{z;z'} * \ln \cZ_L(z',\bz') & = &-\ln \cZ_{L+1}(z,\bz) \\[4pt]
    [\hat{\,\bf L}_z]^k * \ln\cZ_0(z,\bz) & = &\cD_0^k(z,\bz) \\
    \bottomrule
    \end{tabular}
    \caption{Actions of the operators on thermal free energies.}
    \label{tab:my_label_3}
\end{table}

\section{The Thermal Representation of Conformal Ladder Integrals}
\label{sec: correspondence}

\subsection{The correspondence}

Comparing (\ref{ILk_our1}) with (\ref{Zdexpl}) and (\ref{ZL}), the identifications $\ze\leftrightarrow z$ and $g^2\leftrightarrow\a^2$ yield our main result:
\begin{mybox}
\begin{align}
\label{ILk_res1}
\cI_L^k(\ze,\bze)=\frac{\Phi_L^k(\ze,\bze)}{|1-\ze|^{2k}}
\;\longleftrightarrow\;
&\frac{1}{\Gamma(k)L!(z-\bz)^k}\,[-\check{\bf d}_{z;z'}]^L*[\hat{\bf{L}}_{z'}]^k*\ln\cZ_0(z',\bz')\nonumber\\
&\equiv\frac{1}{\Gamma(k)L!(z-\bz)^k}\,\cD_L^k(z,\bz)\,.
\end{align}
\end{mybox}
Equivalently,
\begin{mybox}
\be
\label{ILk_res11}
\Phi_L^k(z,\bz)=\frac{1}{L!}\frac{\cD_L^k(z,\bz)}{\cD_0^k(z,\bz)}\,.
\ee
\end{mybox}
The conformal integrals $\cI_L^k$ are thus built from the parent free energy $\ln\cZ_0$ by successive application of $\check{\bf d}_{z;z'}$ and $\hat{\,\bf L}_z^{(k)}$. Notice that the condition $g^2\ll 1$ maps to $\a^2\ll 1$, corresponding to low temperatures. This suggests, intriguingly,  that the conformal integrals of a perturbative CFT in $D$-dimensions correspond to the low dimension/strong coupling regime of $d$-dimensional thermal field theories. The correspondence between the two sets of variables is summarised in Table \ref{tab:correspondence-CFT}.

\begin{table}[h!]
    \centering
    \begin{tabular}{ccc}
    \toprule
    \textbf{Conformal ladder integrals} && \textbf{Thermal one-point functions} \\
    \midrule
    Dimension $D$ & $D=2k+2$ & Depth $k$ \\
    Loop order $L$ & $d=2L+1$ & Dimension $d$ \\
    Cross-ratios $x_i=(0,1,z,\infty)$ & $z=e^{-\beta m-i\beta\mu}$ & Mass $m$, chemical potential $\mu$ \\
    Coupling $g^2$ & $g^2=\a^2$ & Geometric parameter $\a^2=\tfrac{\ell^2}{4\pi\beta^2}$ \\
    \bottomrule
    \end{tabular}
    \caption{Correspondence between conformal ladder integrals and thermal one-point functions.}
    \label{tab:correspondence-CFT}
\end{table}

\subsection{A familiar-looking differential equation}

Starting from the integral representation
\be
\label{lnZL}
\ln\cZ_L(z,\bz)=-\frac{1}{(L-1)!}\int_0^{|z|}\frac{d|z'|}{|z'|}2\ln|z'|\,
(\ln^2|z'|-\ln^2|z|)^{L-1}\ln\cZ_0(z',\bz')\,,
\ee
and applying the Laplacian (\ref{Laplacian}), using $(z\partial_z+\bz\partial_\bz)\ln\cZ_0=0$ we found in  \cite{Karydas:2025tfs}
\be
\label{DZ01}
\tfrac{1}{4\b^2}\hat{\bf\Delta}_z*\ln\cZ_L(z,\bz)=L\hat{\bf D}_z*\ln\cZ_L(z,\bz)\,.
\ee
Together with the commutator $[\tfrac{1}{4\b^2}\hat{\bf\Delta}_z,\hat{\,\bf L}_z^{(k)}]=-2k\tfrac{|z|^2}{(z-\bz)^2}\hat{\,\bf L}_z^{(1)}$, induction shows that $\cD_L^k(z,\bz)$ satisfies
\begin{mybox}
\be
\label{EqDLk}
\left(\frac{1}{4\b^2}\hat{\bf\Delta}_z-L\hat{\bf D}_z+k(k-1)\frac{|z|^2}{(z-\bz)^2}\right)*\cD_L^k(z,\bz)=0\,.
\ee
\end{mybox}
In terms of $(m,\mu)$ variables this reads
\begin{mybox}
\be
\label{EqDLk1}
\left[\frac{\partial^2}{\partial m^2}+\frac{\partial^2}{\partial\mu^2}
-\frac{2L}{m}\frac{\partial}{\partial m}-\b^2\frac{k(k-1)}{\sin^2(\b\m)}\right]\cD_L^k(m,\m)=0\,.
\ee
\end{mybox}
For $L=0$ this is the equation of motion of a scalar field on Euclidean AdS$_2$ with metric $ds^2=(\sin^2(\b\m))^{-1}(dm^2+d\mu^2)$ and mass $M^2=\b^2k(k-1)$, suggesting that the $\cD_0^k$ are bulk-to-boundary propagators in an AdS$_2$/CFT$_1$ setting. This is reminiscent of a recent observation \cite{Hartnoll:2025hly} on the relation of AdS$_2$/CFT$_1$ to conformal quantum mechanics \cite{Chamon:2011xk}, and this becomes intriguing in view of the connection of the latter to conformal ladder graphs \cite{Isaev:2003tk}.  It would also be very interesting to investigate whether there exists a similar holographic interpretation for the functions $\cD_L^k$ for $L> 0$.

\begin{figure}[h!]
    \centering
    \includegraphics[width=0.8\linewidth]{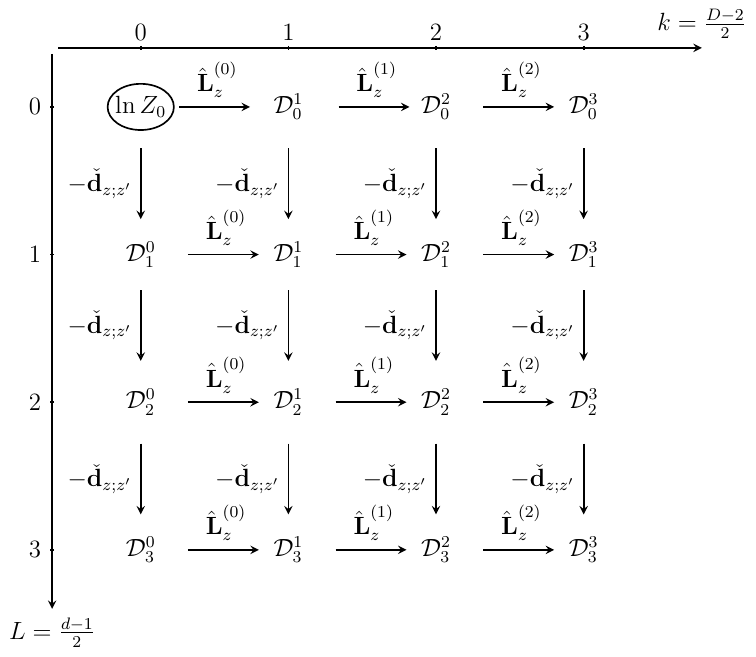}
    \caption{Conformal ladder integrals built from the partition function $\ln \mathcal{Z}_{0}$}
    \label{fig: from HO to conf integrals}
\end{figure}
\section{Applications}
\label{sec: applications}

\subsection{Higher-spin thermal one-point functions as conformal ladder integrals}

The thermal averages of $\cO$ and $\cQ$ in $d=2L+1$ dimensions are
\be
\label{OQ}
\langle\cO(z,\bz)\rangle_L\equiv\beta\cD_{L-1}^0(z,\bz)\,,\qquad
\langle\cQ(z,\bz)\rangle_L\equiv\cD_L^1(z,\bz)\,,
\ee
i.e.\ conformal ladder integrals in $D=2$ and $D=4$ respectively. These appear in the OPE expansion of the thermal two-point function of a free massive complex scalar \cite{Iliesiu:2018fao},
\be
\label{phiphi}
g^{(L)}(r,\theta;0,0)=\frac{1}{r^{2\Delta_\phi}}\!\left[C_\phi^L(1)+\sum_{\{\cO_s\}}a^L_{\cO_s}
\!\left(\frac{r}{\beta}\right)^{\!\!\Delta_{\cO_s}}\!\!C_s^{\nu}(\cos\theta)\right],\quad\nu=L-\tfrac{1}{2}\,.
\ee
At finite mass and imaginary chemical potential, the thermal one-point coefficients become (for $s\geq 0$) \cite{Karydas:2023ufs}
\begin{align}
\label{aOs}
a^{L}_{\cO_s}(z,\bz)&=\frac{\G(L-\frac{1}{2})}{\G(L+s-\frac{1}{2})(4\pi)^L 2^{2s}}
\sum_{n=0}^{L-1+s}\frac{2^n(\beta m)^n(2L-2+s-n)!}{n!\,(L-1+s-n)!}\nonumber\\
&\quad\times\left[Li_{2L-1+s-n}(z)+(-1)^s Li_{2L-1+s-n}(\bar{z})\right]\,.
\end{align}
These coefficients satisfy the following recurrence relation proven in Appendix D of \cite{Karydas:2025tfs}
\be
\label{recursion}
a^L_{\cO_{s+2}}(z,\bz)=\frac{2\pi}{2L-1}a^{L+1}_{\cO_s}(z,\bz)
+\frac{\ln^2|z|}{(2L-1+2s)(2L+1+2s)}a^L_{\cO_s}(z,\bz)\,.
\ee
This shows that all higher-spin thermal one-point functions are determined by the spin-0 and spin-1 coefficients, and hence by conformal ladder integrals in $D=2$ and $D=4$. This is depicted in Fig. 3.
\begin{figure}[h!]
    \centering
    \includegraphics[width=0.8\linewidth]{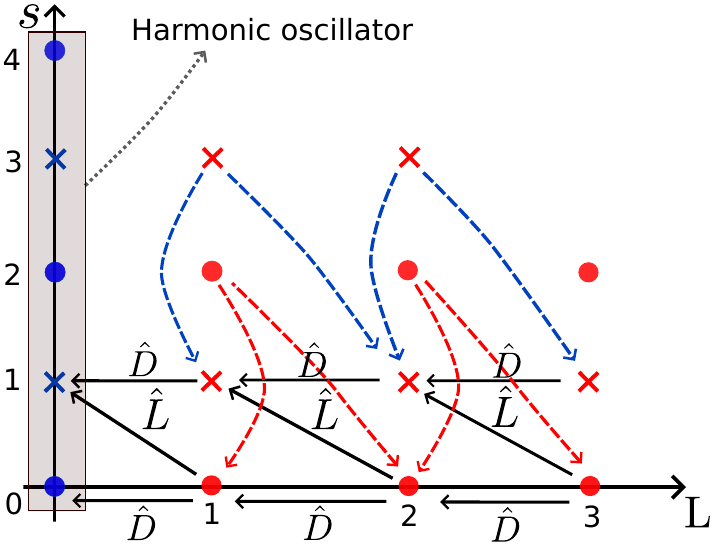}
    \caption{Relationships between the $a_{\mathcal{O}_{s}}$. The dashed lines represent algebraic relations, while the solid lines represent differential ones.}
    \label{fig: relations aL}
\end{figure}

\subsection{The hyper-partition function and the all-loop resummation}

We define the \textit{hyper-partition function}
\be
\label{hyperZ}
\ln\cZ(z,\bz;\a^2)=\sum_{L=0}^\infty\ln\cZ_L(z,\bz;\a^2)\,,
\ee
for which (\ref{Oaverage}) implies the first-order equation
\be
\label{dhyperF1}
\left(\frac{\partial}{\partial m}+2\a^2\b^2 m\right)\ln\cZ(m,\m;\a^2)
=-\b\frac{\sinh\b m}{\cosh\b m-\cos\b\m}\,.
\ee
For the all-loop resummation of conformal ladder integrals, we form the Borel sum of the free energies via the analytic continuation $\a\to ig$:
\begin{align}
\label{dhyperfF6}
\sum_{L=0}^{\infty}\frac{1}{L!}\ln\cZ_L(z,\bz;-g^2)&=\ln\cZ_0(m,\m)
+\left[J_0(2g\b\sqrt{\omega^2-m^2})\ln\cZ_0(\omega,\m)\right]_m^\infty\nonumber\\
&\quad+\b\int_m^\infty\!d\omega\,J_0(2g\b\sqrt{\omega^2-m^2})\frac{\sinh(\b\omega)}{\cosh(\b\omega)-\cos(\b\m)}\,.
\end{align}
The all-loop resummation (\ref{ILk_res1}) then gives
\be
\label{sum1}
\cI^k(z,\bz;g^2)=\sum_{L=0}^\infty(-g^2)^L\cI_L^k(z,\bz)
=\frac{1}{\Gamma(k)(z-\bz)^k}\left[\hat{\,\bf L}_z\right]^k*\sum_{L=0}^\infty\frac{1}{L!}\ln\cZ_L(z,\bz;-g^2)\,,
\ee
and for $k=1$ ($D=4$) one obtains
\begin{align}
\cI^1(z,\bz;g^2)
=\frac{1}{2|z|}\int_m^\infty\!d\omega\,J_0(2g\b\sqrt{\omega^2-m^2})
\frac{\sinh(\b\omega)}{\left(\cosh(\b\omega)-\cos(\b\m)\right)^2}\,,
\end{align}
matching a famous old result by Broadhurst et al.\ \cite{Broadhurst:1993ib,Broadhurst:2010ds}.

\section{Summary and Outlook}
\label{sec: conclusion}

We have established a \textit{thermal representation for conformal ladder integrals}: an explicit relationship between $L$-loop conformal ladder integrals in even $D$ dimensions and thermal averages in free massive complex scalar theories: the loop order $L$ maps to spacetime dimension $d=2L+1$ and the dimension where the integrals are performed $D$ maps to a depth parameter $k$ via $D=2k+2$. All integrals are constructed from the free energy of a parent quantum mechanical system of harmonic oscillators twisted by an imaginary chemical potential. Using this representation we uncovered a novel second-order differential equation satisfied by all conformal ladder integrals and derived, in a simple way, an all-loop resummation formula.

Several open directions are noteworthy:

\textbf{Relation to the AGT conjecture.} The mapping of spacetime cross-ratios to thermal variables resembles the AGT conjecture \cite{Alday:2009aq}, but our construction does not rely on supersymmetry or complex geometry; understanding better this relationship could be fruitful.

\textbf{Large-charge calculations.} The all-loop resummation has recently appeared in large-charge calculations in the $O(N)$ vector model \cite{Giombi:2020enj} and ${\cal N}=4$ SYM \cite{Caetano:2023zwe,Brown:2025cbz}. Our thermal representation associates these with a sum of free energies reminiscent of a genus expansion of string amplitudes.

\textbf{Thermal bootstrap.} Thermal one-point functions are related to conformal integrals, so symmetry properties of the latter—including KMS constraints—may be studied through our differential equation \cite{Marchetto:2023xap,Barrat:2025wbi,Barrat:2025nvu,Buric:2025anb,Buric:2025uqt}.

\textbf{Modular forms and string amplitudes.} Our free energies satisfy equations and recurrence relations similar to those for modular forms \cite{Alessio:2021krn,Aggarwal:2024axv}, opening the possibility of connecting recent progress in string theory amplitudes \cite{Dorigoni:2021ngn,Dorigoni:2021jfr} to thermal CFTs.

\textbf{CFTs on  $S^1\times S^{d-1}$.} Thermal one-point functions arise as large-volume limits of the one-point functions on $S^1\times S^{d-1}$  \cite{David:2024pir,Alkalaev:2024jxh,Buric:2024kxo,Buric:2025uqt}; it would be interesting to study how the differential equations and other relations of the conformal ladder integrals arise from the thermal conformal blocks.

\textbf{Conformal higher-derivative theories.} Conformal ladder integrals in higher-derivative CFTs \cite{Brust:2016gjy} have been recently studied in \cite{Giombi:2022gjj} and their thermal properties in \cite{Benedetti:2023pbt}. Our thermal representation can be applied in the study of these models, which are thought to describe long-range critical systems.

\textbf{Integrability properties of multiloop Feynman graphs.}\footnote{It is hard to do justice to the wealth of important works in this direction which include among others \cite{Basso:2017jwq,Vanhove:2018elu, Charlton:2021uhu,Travaglini:2022uwo}.} Recently our $\hat{\bf D}_z$ and $\hat{\,\bf L}_z$ operators were used in \cite{Loebbert:2024fsj} to show that ladder integrals obey Toda-like equations, and also in \cite{Dixon:2025zwj}  to demonstrate an antipodal symmetry of conformal integrals. One may also try to extend the thermal representation to pentaladders \cite{Caron-Huot:2018dsv}. We believe that there is more to be uncovered in these directions; see e.g.\ \cite{He:2025lzd} for a recent work.

\textbf{Quantum gravity connection.} The parent quantum mechanical system (\ref{BasicHamiltonian}) may be thought of as a primon gas whose partition function gives Wheeler-DeWitt eigenfunctions describing quantum gravity near a singularity \cite{Hartnoll:2025hly,DeClerck:2025mem}, satisfying differential equations similar to (\ref{EqDLk1}).

\section*{Acknowledgments}

A.C.P.\ wishes to acknowledge enlightening discussions and correspondence with G.~Barnich, M.~Berg, J.~David, S.~Giombi, V.~Kazakov, A.~Kleinschmidt, E.~Marchetto, A.~Miscioscia, N.~Obers, E.~Pomoni, A.~Santambrogio, L.~Shumilov, C.~Wen and K.~Zarembo. He thanks the Simons Center for Geometry and Physics at Stony Brook for hospitality during `Black Hole Physics from Strongly Coupled Thermal Dynamics'. The work of M.V.\ was supported by the F.R.S.--FNRS under Grant No.\ T.0047.24.

\appendix

\end{document}